\documentclass[cameraready]{Interspeech}
\title{SimulU: Training-free Policy for Long-form Simultaneous\\Speech-to-Speech Translation}

\author[affiliation={1}, orcid=0000-0003-0831-4867]{Amirbek}{Djanibekov}
\author[affiliation={2}, orcid=0000-0001-7480-2231]{Luisa}{Bentivogli}
\author[affiliation={2}, orcid=0000-0002-8811-4330]{Matteo}{Negri}
\author[affiliation={2}, orcid=0000-0002-4494-8886]{Sara}{Papi}
\address{
    $^1$MBZUAI, United Arab Emirates \\
    $^2$FBK, Italy
}

\email{amirbek.djanibekov@mbzuai.ac.ae, spapi@fbk.eu}


\usepackage{comment}
\usepackage{lipsum}
\usepackage{pgfplots}
\usepackage{pgfplotstable}

\usepackage{multicol}
\usepackage{multirow}
\usepackage{soul}
\usepackage{colortbl}

\pgfplotsset{compat=1.18}

\usepackage{xcolor}

\definecolor{colSimulU}{HTML}{1f77b4}      % blue
\definecolor{colSimulUmono}{HTML}{8C564B}      % brown
\definecolor{colStreamAttXTTS}{HTML}{ff7f0e}     % orange
\definecolor{colStreamAttSeamless}{HTML}{2ca02c} % green
\definecolor{colLASeamless}{HTML}{d62728} % red
\definecolor{colLAXTTS}{HTML}{9467bd}     % purple

\begin{document}

\maketitle

\begin{abstract}
Simultaneous speech-to-speech translation (SimulS2S) is essential for real-time multilingual communication, with increasing integration into meeting and streaming platforms. Despite this, SimulS2S remains underexplored in research, where current solutions often rely on resource-intensive training procedures and operate on short-form, pre-segmented utterances, failing to generalize to continuous speech. To bridge this gap, we propose SimulU, the first training-free policy for long-form SimulS2S. SimulU adopts history management and speech output selection strategies that exploit cross-attention in pre-trained end-to-end models to regulate both input history and output generation. Evaluations on MuST-C across 8 languages show that SimulU achieves a better or comparable quality-latency trade-off against strong cascaded models. By eliminating the need for ad-hoc training, SimulU offers a promising path to end-to-end SimulS2S in realistic, long-form scenarios.
\end{abstract}

\section{Introduction}
Speech-to-speech (S2S) technology represents a high-potential direction for enhancing natural and seamless human-computer interaction~\cite{munteanu2013we,munteanu2017speech}, enabling end-to-end spoken communication across languages and modalities \cite{jia19_interspeech}.
Within this context, simultaneous translation (a.k.a. simultaneous interpretation) extends the S2S paradigm by requiring a system to generate translated speech incrementally as the input stream is received.
This setup mandates a real-time decision-making policy to balance the trade-off between reading new input and writing output based on partial information, often under acoustic and linguistic uncertainty.
The optimal design of such a \textit{simultaneous policy}
is further complicated for long-form inputs, where simultaneous S2S translation (SimulS2ST) operates on continuous, unsegmented speech streams rather than pre-segmented utterances.

The SimulS2ST policy is usually learned during complex training pipelines.
For instance,~\cite{zhang2024streamspeech} jointly optimizes four distinct objectives, while~\cite{deng-etal-2025-simuls2s} adopts a two-stage procedure incorporating a large language model~\cite{grattafiori2024llama3herdmodels}.
More recent approaches, such as~\cite{cheng2025seedliveinterpret20endtoend} and~\cite{labiausse2026simultaneousspeechtospeechtranslationaligned}, further introduce reinforcement learning to refine policy learning.
Additionally, the training process is compounded by the need for large-scale speech data~\cite{labiausse2025highfidelity}. 
The limited availability of word-level aligned corpora necessitates the use of synthetic datasets, where alignments are automatically generated through hand-crafted heuristics~\cite{labiausse2025highfidelity, labiausse2026simultaneousspeechtospeechtranslationaligned} (e.g., natural pauses).
Standard approaches typically use cascaded pipelines that include speech-to-text translation (S2TT) and text-to-speech (TTS) components~\cite{lavie1997janus,nakamura2006atr}. 
However, cascaded systems have several problems. 
First, they are subject to compounding errors due to combining separately trained models \cite{sperber-paulik-2020-speech}. 
Second, as input speech goes through the text bottleneck, the non-linguistic information it carries (e.g., speaker identity, prosody) is lost and cannot be transferred to the output speech \cite{tsiamas-etal-2024-speech}. 
Finally, cascaded pipelines are inherently disadvantaged in latency-critical settings, as each component must complete its processing before the next can begin, making such systems less suited for simultaneous translation.
In addition, most existing works evaluate SimulS2ST in a short-form setting \cite{10889740,deng-etal-2025-simuls2s}, 
relying on test sets inherited from offline scenarios where input audio is pre-segmented (often manually) into fixed-length chunks (typically up to 30 seconds). This setup constrains systems to operate within predefined boundaries, limiting their ability to handle long-form, continuous speech and diverging from more realistic deployment conditions \cite{papi-etal-2025-real}.

To fill these gaps, we propose \textbf{SimulU}, the first training-free simultaneous policy for long-form end-to-end speech-to-speech translation. 
Building on the recent success of exploiting attention scores for guiding simultaneous S2TT inference \cite{papi-etal-2024-streamatt}, we propose leveraging cross-attention not only to decide what and when to emit a \textit{partial} spoken translation, but also to determine which contextual history--both from the received speech input and the generated output--to retain, thereby enabling long-form speech generation. 
These decisions are made solely based on the internal knowledge of pre-existing models that natively incorporate attention mechanisms, without requiring retraining or adaptation.
SimulU eliminates the need for costly ad-hoc training procedures while addressing the read/write decision problem directly in an end-to-end setting.
We showcase our proposed policy by applying it to a strong offline pretrained model, SeamlessM4T~\cite{communication2023seamlessm4tmassivelymultilingual}, and comparing it against strong baselines built on state-of-the-art ASR, S2TT, and TTS components across all 8 languages of MuST-C v1.0~\cite{di-gangi-etal-2019-must}. 
Our method achieves the best quality-latency trade-off in most settings, providing the first promising step in the training-free policy research for long-form speech-to-speech translation.

\section{Methodology}
\begin{figure}[t]
    \centering
    \includegraphics[width=\linewidth]{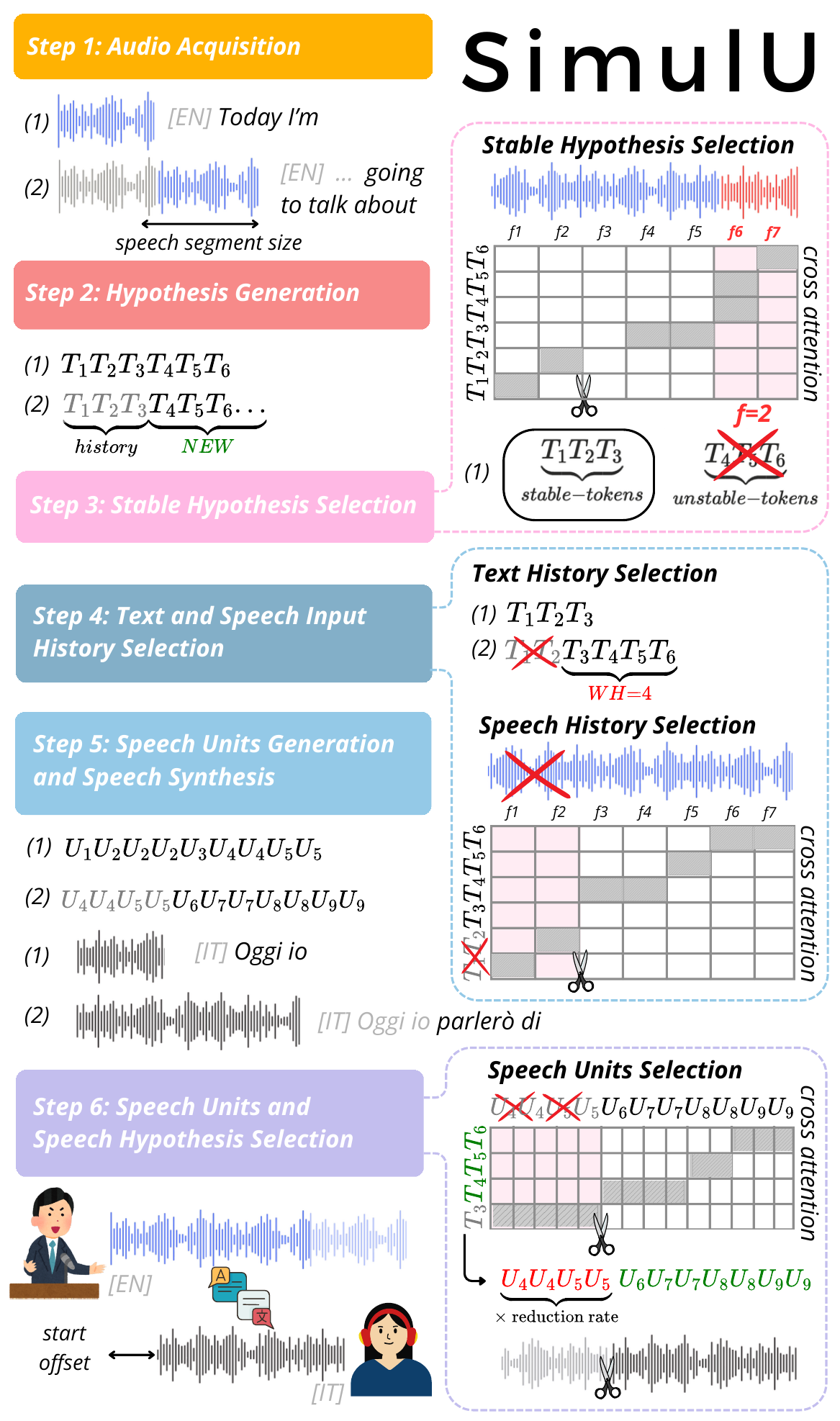}
    \caption{SimulU Overview. The incoming speech input, here in English, is represented by the \textcolor{blue}{blue waveform}, while the speech output, here in Italian, is represented by the \textbf{black waveform}.}
    \label{fig:system_overview}
\end{figure}

SimulU is a simultaneous policy that implements history management and speech output selection for direct SimulS2ST based on internal knowledge of the system, in particular, the cross-attention scores. The policy is applied to the SeamlessM4T S2ST model, which employs a speech-to-text module, a text-to-unit module, and a vocoder (more details on the SimulU backbone can be found in Section \ref{subsec:models}) that were jointly trained for \textit{offline} end-to-end S2ST.
 
Without requiring additional training or adaptation, SimulU repurposes attention-based offline models (here SeamlessM4T) for simultaneous generation, a process frequently referred to as \textit{onlinization} \cite{machacek-polak-2025-simultaneous,polak-etal-2022-cuni}.
Its workflow follows a six-step policy, illustrated in Figure \ref{fig:system_overview} and detailed below:

\begin{enumerate}
    \item \textbf{Audio Acquisition}: the incoming speech input, divided in chunks with size \textit{speech segment size}, is incrementally added to the \textbf{\textit{speech history}}.
    \item \textbf{Hypothesis Generation}: the content of the \textbf{\textit{speech history}} is provided to the speech-to-text module to generate intermediate textual representations.
    \item \textbf{Stable Hypothesis Selection}: the intermediate textual representation is selected to be given as context to Step 5 based on speech-text cross-attention scores following \cite{papi23_interspeech}, which stops the hypothesis emission as soon as a token is aligned (i.e., having the maximum attention score) with one of the unstable audio frames (i.e., the lastly received speech frames); the unstable audio frames, hereinafter \textit{cut-off frame} defined by the $f$ hyperparameter (and equal to $2$ in Figure \ref{fig:system_overview}), control the latency of the system.
    \item \textbf{Text and Speech Input History Selection}: to allow for long-form speech processing, both \textbf{\textit{text history}} and \textit{\textbf{speech history}} (at input only) have to be selected to maintain a context that is manageable by the model.
    %; we 
   To this aim, we 
   preserve a fixed number of words, hereinafter $WH$ (equal to $4$ in Figure \ref{fig:system_overview}) in the \textbf{\textit{text history}}, following \cite{papi-etal-2024-streamatt}, and exploit speech-text cross-attention scores again to select the audio frames corresponding to the \textit{discarded} \textbf{\textit{text history}} (equal to $2$ in Figure \ref{fig:system_overview}), which are removed from the \textit{\textbf{speech history}} to always keep the text and speech content aligned.
    \item \textbf{Speech Units Generation and Speech Synthesis}: the intermediate textual representation is provided to the text-to-unit module and, consequently, to the vocoder to generate the output speech; the whole \textbf{\textit{text history}} is provided in this phase, as we found it dramatically improves the synthesized speech.
    \item \textbf{Speech Units and Speech Hypothesis Selection}: the text-unit cross-attention is leveraged in the final step, which is in charge of selecting the part of the output speech that corresponds to the newly generated hypothesis; by taking the maximum attention scores (similar to Step 3 and 4), the intermediate textual representation is aligned with the corresponding units, and the units assigned to the text history but the new hypothesis ($T_3$ in Figure \ref{fig:system_comparison}) are discarded; the number of discarded units is then multiplied by the \textit{reduction rate} of SeamlessM4T (i.e., $320$) to retrieve the part of the synthesized waveform to be cut from the output speech. The resulting partial waveform is emitted. 
\end{enumerate}

\section{Experimental Settings}

\begin{table*}[h]
\centering
\resizebox{\linewidth}{!}{
    \begin{tabular}{cccccccccccccccccccc}
    \toprule
    
     & 
    \multicolumn{3}{c}{\textit{Word History = 5}} & \multicolumn{4}{c}{\textit{Word History = 10}} & \multicolumn{4}{c}{\textit{Word History = 15}} & \multicolumn{4}{c}{\textit{Word History = 20}} & \multicolumn{4}{c}{\textit{Word History = 25}} \\
     \cmidrule{2-4} \cmidrule{6-8} \cmidrule{10-12} \cmidrule{14-16} \cmidrule{18-20}

     & ASR & Start & End & & 
    ASR & Start & End & & 
    ASR & Start & End & & ASR & Start & End & & ASR & Start & End\\
        
     & BLEU & Offset & Offset & & 
    BLEU & Offset & Offset & & 
    BLEU & Offset & Offset & & BLEU & Offset & Offset & & BLEU & Offset & Offset \\
    \midrule
    2  & 17.97 & 0.90 & 77.24 & & 18.56 & 0.90 & 128.68 & & 19.38 & 0.90 & 161.47 & & 16.60 & 0.90 & 187.27 & & 13.63 & 0.90 & 183.51 \\
    4  & 16.93 & 1.36 & 51.18 & & 23.04 & 1.24 & 147.60 & & 18.65 & 1.36 & 151.84 & & 18.15 & 1.36 & 240.88 & & 12.73 & 1.36 & 202.10 \\
    6  & 16.54 & 1.72 & 55.95 & & 18.85 & 1.72 & 131.39 & & 20.32 & 1.72 & 179.22 & & 17.87 & 1.72 & 256.14 & & 11.37 & 1.72 & 184.63 \\
    8  & 18.38 & 1.77 & 46.78 & & 17.64 & 1.77 & 121.12 & & 20.12 & 1.77 & 189.09 & & 18.25 & 1.77 & 263.51 & & 11.37 & 1.77 & 173.28 \\
    \bottomrule \\
    \end{tabular}
}
    \caption{ASR-BLEU scores for different Word History ($WH$) configurations (5-25) across combinations of cut-off frame $f$ (2, 4, 6, 8).}
    \label{tab:devset_word_history}
\end{table*}

\subsection{Data}
To be comparable with previous S2TT works~\cite{papi-etal-2024-streamatt,ouyang-etal-2025-infinisst}, we evaluated on all languages of
MuST-C v1.0~\cite{di-gangi-etal-2019-must}, English (en) to Dutch (nl), French (fr), German (de), Italian (it), Portuguese (pt), Russian (ru), Romanian (ro), Spanish (es). 
We simulate streaming conditions by providing the entire TED Talks from the MuST-C dev and tst-COMMON set as input. The average durations for development and test sets are 948 seconds (15.7 minutes) and 650 seconds (10.8 minutes), respectively.

\subsection{Metrics}
We follow the settings from the IWSLT 2023 evaluation campaign on simultaneos speech-to-speech translation~\cite{agrawal-etal-2023-findings}, which measures both quality and latency using SimulEval~\cite{simuleval2020}.
For quality, BLEU score~\cite{papineni2002bleu} is computed on the transcripts obtained with the Canary~\cite{sekoyan2025canary} model on the translated output speech.
We use Canary over Whisper~\cite{radford2023robust} because we empirically found that Canary handles long speech better than Whisper.
Additionally, we applied the Whisper text normalizer before computing the BLEU score.
For latency, we report StartOffset following IWSLT 2023 settings (negative results are excluded, as they are due to misalignment). 
It is measured in seconds and represents the minimum amount of source input that must be observed before the system emits the first target output. A graphical representation is provided in Figure~\ref{fig:system_overview}.

\subsection{Models}
\label{subsec:models}

\textbf{SimulU Backbone.}
We employ the SeamlessM4T model (\texttt{SeamlessM4T-medium-v1}\footnote{\url{https://huggingface.co/facebook/hf-seamless-m4t-medium}}), a series of transformer encoder-decoder architecture~\cite{communication2023seamlessm4tmassivelymultilingual} for translation. 
The S2T model uses 2 frame stacking, 256k word-piece tokenization, and supports around 100 languages, resulting in $\sim$1B total parameters.
The model takes a mel-spectrogram with a hop size of 160 as input and produces tokenized text as output.
The speech encoder follows the w2v-BERT~\cite{chung2021w2v} framework and consists of 12 Conformer layers, totaling approximately 300 million parameters.
For the text decoder, SeamlessM4T adopts the NLLB~\cite{costa2022no} decoder trained on around 100 languages, rather than the 200 languages used in the original NLLB study.
Additionally, the NLLB text encoder is employed for knowledge distillation, aligning the speech encoder’s representations with the text embedding space using the SONAR~\cite{duquenne2023sonar} alignment score.
The TTS employs a transformer encoder–decoder (170M parameters) that produces speech units at 50Hz.
It outputs discrete speech units derived from XLS-R-1B 35th layer representations~\cite{babu2022xls} using (k)-means clustering, and uses a specially designed interleaving between speech encoded representation and decoded translation text.
Finally, the generated speech units are sent to a unit-vocoder, which is a multilingual HiFi-GAN~\cite{kong2020hifi} unit that synthesizes speech from those units.

To determine optimal parameter setting for the word history of SimulU, we did preliminary experiments on the MuST-C dev set in the en-de direction. 
The results are shown in Table~\ref{tab:devset_word_history}. 
Overall, the SimulU configuration using a word history of $10$ yields results that are better than or on par with the others, while keeping the history relatively short, and therefore used throughout the rest of the paper.

\noindent \textbf{Baselines.}
We developed four strong cascade approaches based on state-of-the-art training-free S2TT policies, StreamAtt \cite{papi-etal-2024-streamatt} and LocalAgreement (LA~\cite{liu2020low}), which are then coupled with existing TTS models for the speech generation part:
\begin{itemize}
    \item \textbf{StreamAtt+SeamTTS} is based on StreamAtt \cite{papi-etal-2024-streamatt}, which enables long-form S2TT by leveraging cross-attention between speech and generated text to both guide simultaneous inference and history management; the partial generated text is then given to the SeamlessM4T TTS model (Seam.TTS), which is based on the unit-generation architecture of UnitY~\cite{inaguma-etal-2023-unity} for the speech generation. This baseline directly compares the cascaded approach to the end-to-end SimulU approach within the same model, highlighting the performance difference under the same data and architecture.
    \item \textbf{StreamAtt+XTTS-v2} is a cascade approach made of the state-of-the-art S2TT policy, StreamAtt, and the strongest multilingual TTS system that supports 17 languages, except Romanian, achieving the best result on the TTS Arena~\footnote{\url{https://huggingface.co/spaces/TTS-AGI/TTS-Arena}} (best multilingual TTS results after monolingual-English KokoroTTS~\cite{kokoro} and Fish Speech~\cite{fish-speech-v1.4}). This system is considered an upperbound of the TTS performance, as shown in Table \ref{tab:wer_cer_tts_systems}, where XTTSv2 achieves WER and CER scores from 4 to 10 times lower (hence, better) than Seam.TTS, while being comparable or slightly worse regarding naturalness (UTMOS~\cite{saeki2022utmos} from VoiceMOS challenge).
    \item \textbf{LA+Seam.TTS} is a baseline system derived from the IWSLT 2025 simultaneous Speech-to-Text evaluation campaign \cite{agostinelli-etal-2025-findings}. It uses Local Agreement (LA) for the STT part, which compares consecutive chunk outputs and emits only the longest common prefix between the current chunk and the previous chunk's output, ensuring stable hypothesis emission. To allow for long-form processing, silero VAD \cite{SileroVAD} is used to segment the continuous stream of audio into shorter segments of about 15-30s, with a maximum unvoiced interval of 20 seconds and a voice threshold of 0.1, suitable for standard S2TT model, and the memory is reset between segments. In this version, Seam.TTS is used as the TTS component.
    \item \textbf{LA+XTTS-v2}, similar to \textbf{LA+Seam.TTS}, couples the LA policy but replaces Seam.TTS with XTTS-v2 model for the TTS component.
\end{itemize}

For the StreamAtt-based cascades, the latency is controlled by the cut-off frame (spanning 2, 4, 6, and 8), while for the LA-based cascades, by the speech segment size (spanning 0.5, 1.0, 1.5, and 2.0 seconds). Default decoding parameters are used for all models (e.g. num. beams for Seam.TTS is 1).

\section{Results}

\pgfplotstableread[row sep=\\]{
ASR-BLEU	StartOffset    Frames\\
19.536     0.786   2\\
20.436     1.008   4\\
20.771     1.175   6\\
19.859     1.471   8\\
}\endeStreamAttXTTS
\pgfplotstableread[row sep=\\]{
ASR-BLEU	StartOffset    Frames\\
30.079     1.101   2\\
31.017     1.305   4\\
31.295     1.694   6\\
30.082     2.138   8\\
}\enfrStreamAttXTTS
\pgfplotstableread[row sep=\\]{
ASR-BLEU	StartOffset    Frames\\
20.562     0.642   2\\
20.447     1.012   4\\
20.481     1.364   6\\
21.521     1.698   8\\
}\enitStreamAttXTTS
\pgfplotstableread[row sep=\\]{
ASR-BLEU	StartOffset    Frames\\
% 23.83      11.189   2\\
25.341     1.157   4\\
25.188     1.453   6\\
25.891     1.842   8\\
}\enesStreamAttXTTS
\pgfplotstableread[row sep=\\]{
ASR-BLEU	StartOffset    Frames\\
26.342     0.029   2\\
27.568     0.270  4\\
27.89      0.584  6\\
27.938     0.899  8\\
}\ennlStreamAttXTTS
\pgfplotstableread[row sep=\\]{
ASR-BLEU	StartOffset    Frames\\
% 26.165      0.044  2\\
26.915      0.325  4\\
26.738      0.714  6\\
27.169      0.992  8\\
}\enptStreamAttXTTS
\pgfplotstableread[row sep=\\]{
ASR-BLEU	StartOffset    Frames\\
14.262     1.008  2\\
14.723     1.379  4\\
14.817     1.749  6\\
15.197     2.082  8\\
}\enruStreamAttXTTS

%%%%%%%%%%%%%%%%%%%%%%%%%%%%%%%%%%%%%%%%%%%%%%%%%%%%%%%%%%%%%%%%%%

\pgfplotstableread[row sep=\\]{
ASR-BLEU    StartOffset	Frames\\
20.602      1.120	  2\\
20.509      1.249      4\\
19.967      1.453	  6\\
18.745      1.897	  8\\
}\endeSimulU
\pgfplotstableread[row sep=\\]{
ASR-BLEU    StartOffset    Frames\\
33.256	1.323    2\\
33.898  1.564    4\\
34.556	2.008    6\\
34.405	2.490    8\\
}\enfrSimulU
\pgfplotstableread[row sep=\\]{
ASR-BLEU    StartOffset    Frames\\
23.592  0.827        2\\
24.019  1.216        4\\
24.494  1.679        6\\
24.072  2.050        8\\
}\enitSimulU
\pgfplotstableread[row sep=\\]{
ASR-BLEU	StartOffset	Frames\\
% 27.923	 11.898   2\\
28.639   1.305    4\\
29.120	 1.508   6\\
28.670	 2.045   8\\
}\enesSimulU
\pgfplotstableread[row sep=\\]{
ASR-BLEU	StartOffset	Frames\\
26.165	0.195    2\\
26.387  0.455    4\\
26.472	0.936    6\\
26.585	1.177    8\\
}\ennlSimulU
\pgfplotstableread[row sep=\\]{
ASR-BLEU	StartOffset	Frames\\
26.507	0.103        2\\
26.343  0.529        4\\
26.872	1.121        6\\
26.430	1.492        8\\
}\enptSimulU
\pgfplotstableread[row sep=\\]{
ASR-BLEU	StartOffset	Frames\\
13.068	1.490    2\\
12.91   1.860    4\\
12.356	2.101    6\\
12.492	2.582    8\\
}\enruSimulU
\pgfplotstableread[row sep=\\]{
ASR-BLEU	StartOffset	Frames\\
21.305	1.120        2\\
21.909  1.527        4\\
21.628	1.971        6\\
21.720	2.342        8\\
}\enroSimulU

%%%%%%%%%%%%%%%%%%%%%%%%%%%%%%%%%%%%%%%%%%%%%%%%%%%%%%%%%%%%%%%%%%
\pgfplotstableread[row sep=\\]{
ASR-BLEU    StartOffset	Stepsize\\
% 18.669     0.555	  0.5\\
18.065     1.055	  1.0\\
17.948     1.795	  1.5\\
17.394     2.907	  2.0\\
}\endeLAXTTS
\pgfplotstableread[row sep=\\]{
ASR-BLEU    StartOffset	Stepsize\\
% 27.989     0.472	  0.5\\
27.31     1.582	  1.0\\
27.559     2.027	  1.5\\
27.047     2.907	  2.0\\
}\enfrLAXTTS
\pgfplotstableread[row sep=\\]{
ASR-BLEU    StartOffset	Stepsize\\
% 19.863     0.524	  0.5\\
20.292     1.587	  1.0\\
20.148     2.337	  1.5\\
19.365     3.392	  2.0\\
}\enitLAXTTS
\pgfplotstableread[row sep=\\]{
ASR-BLEU    StartOffset	Stepsize\\
% 25.12     0.324	  0.5\\
25.617     1.184	  1.0\\
25.461     1.944	  1.5\\
24.919     3.110	  2.0\\
}\enesLAXTTS
\pgfplotstableread[row sep=\\]{
ASR-BLEU    StartOffset	Stepsize\\
% 21.637    1.618	  0.5\\
% 21.968    0.211	  1.0\\
21.408    0.547	  1.5\\
21.169    2.473	  2.0\\
}\enptLAXTTS
\pgfplotstableread[row sep=\\]{
ASR-BLEU    StartOffset	Stepsize\\
% 23.267     1.387	  0.5\\
23.43     0.334	  1.0\\
22.878      0.566	  1.5\\
21.914     2.158	  2.0\\
}\ennlLAXTTS
\pgfplotstableread[row sep=\\]{
ASR-BLEU    StartOffset	Stepsize\\
% 13.254     0.222	  0.5\\
13.243     1.258	  1.0\\
13.211     1.601	  1.5\\
13.18     3.166	  2.0\\
}\enruLAXTTS

%%%%%%%%%%%%%%%%%%%%%%%%%%%%%%%%%%%%%%%%%%%%%%%%%%%%%%%%%%%%%%%%%%

\pgfplotstableread[row sep=\\]{
ASR-BLEU    StartOffset	Stepsize\\
% 8.21      0.287	  0.5\\
9.37      0.277	  1.0\\
9.92      0.897	  1.5\\
9.598     1.962	  2.0\\
}\endeLASeamless
\pgfplotstableread[row sep=\\]{
ASR-BLEU    StartOffset	Stepsize\\
% 19.059     0.833	  0.5\\
21.713     0.397	  1.0\\
22.545     0.786	  1.5\\
22.257     1.768	  2.0\\
}\enfrLASeamless
\pgfplotstableread[row sep=\\]{
ASR-BLEU    StartOffset	Stepsize\\
% 11.873      0.709	  0.5\\
13.8        0.411	  1.0\\
14.408      0.855	  1.5\\
14.551      2.059	  2.0\\
}\enitLASeamless
\pgfplotstableread[row sep=\\]{
ASR-BLEU    StartOffset	Stepsize\\
% 12.273      1.313	  0.5\\
% 15.031      0.535	  1.0\\
15.102      0.010  1.5\\
15.273      1.279	  2.0\\
}\enptLASeamless
\pgfplotstableread[row sep=\\]{
ASR-BLEU    StartOffset	Stepsize\\
% 11.551    1.396	  0.5\\
% 14.995    0.489  1.0\\
15.533    0.223  1.5\\
15.575     1.057	  2.0\\
}\ennlLASeamless
\pgfplotstableread[row sep=\\]{
ASR-BLEU    StartOffset	Stepsize\\
% 14.147     0.611  0.5\\
17.747     0.712	  1.0\\
18.706     1.397	  1.5\\
18.965     2.388	  2.0\\
}\enesLASeamless
\pgfplotstableread[row sep=\\]{
ASR-BLEU    StartOffset	Stepsize\\
% 6.466     0.491	  0.5\\
7.444     0.592	  1.0\\
7.742     1.008	  1.5\\
7.61      1.712	  2.0\\
}\enruLASeamless
\pgfplotstableread[row sep=\\]{
ASR-BLEU    StartOffset	Stepsize\\
% 9.621      0.491  0.5\\
11.831     0.462	  1.0\\
12.271     1.240	  1.5\\
11.898     2.166	  2.0\\
}\enroLASeamless

%%%%%%%%%%%%%%%%%%%%%%%%%%%%%%%%%%%%%%%%%%%%%%%%%%%%%%%%%%%%%%%%%%

\pgfplotstableread[row sep=\\]{
ASR-BLEU    StartOffset	Frames\\
2.676     0.934	  2\\
0.695     1.268	  4\\
19.967    1.268	  6\\
18.745    1.897   8\\
}\endeStreamAttSeamless
\pgfplotstableread[row sep=\\]{
ASR-BLEU    StartOffset	Frames\\
11.612  1.286   2\\
9.846   1.508   4\\
8.869   1.916   6\\
8.03    2.527   8\\
}\enfrStreamAttSeamless
\pgfplotstableread[row sep=\\]{
ASR-BLEU    StartOffset	Frames\\
6.932     0.735	  2\\
5.512     1.179	  4\\
4.486     1.494	  6\\
4.158     1.920	  8\\
}\enitStreamAttSeamless
\pgfplotstableread[row sep=\\]{
ASR-BLEU    StartOffset	Frames\\
% 7.215     0.044	  2\\
5.911     0.418	  4\\
5.00      1.010	  6\\
4.479     1.492	  8\\
}\enptStreamAttSeamless
\pgfplotstableread[row sep=\\]{
ASR-BLEU    StartOffset	Frames\\
7.117      0.066	  2\\
4.688      0.344	  4\\
4.231      0.714	  6\\
3.467      1.029	  8\\
}\ennlStreamAttSeamless
\pgfplotstableread[row sep=\\]{
ASR-BLEU    StartOffset	Stepsize\\
%8.702     11.898	  0.5\\
7.748     1.157	  1.0\\
6.937     1.471	  1.5\\
6.452     2.045	  2.0\\
}\enesStreamAttSeamless
\pgfplotstableread[row sep=\\]{
ASR-BLEU    StartOffset	Frames\\
5.21     1.305	  2\\
4.004    1.768	  4\\
3.412    2.064	  6\\
3.159    2.527	  8\\
}\enruStreamAttSeamless
\pgfplotstableread[row sep=\\]{
ASR-BLEU    StartOffset	Frames\\
7.52      1.101	  2\\
5.985     1.564	  4\\
5.415     2.027	  6\\
5.203     2.268	  8\\
}\enroStreamAttSeamless

%%%%%%%%%%%%%%%%%%%%%%%%%%%%%%%%%%%%%%%%%%%%%%%%%%%%%%%%%%%%%%%%%%

% % \pgfplotstableread[row sep=\\]{
% % ASR-BLEU    StartOffset	Frames\\
% % 20.602      1.120	  2\\
% % 20.509      1.249      4\\
% % 19.967      1.453	  6\\
% % 18.745      1.897	  8\\
% % }\endeSimulUmono
% \pgfplotstableread[row sep=\\]{
% ASR-BLEU    StartOffset    Frames\\
% 33.256	1.323    2\\
% 33.898  1.564    4\\
% 34.556	2.008    6\\
% 34.405	2.490    8\\
% }\enfrSimulUmono
% \pgfplotstableread[row sep=\\]{
% ASR-BLEU    StartOffset    Frames\\
% 23.592  0.827        2\\
% 24.019  1.216        4\\
% 24.494  1.679        6\\
% 24.072  2.050        8\\
% }\enitSimulUmono
% \pgfplotstableread[row sep=\\]{
% ASR-BLEU	StartOffset	Frames\\
% 27.923	 1.1898   2\\
% 28.639   1.305    4\\
% 29.120	 1.508   6\\
% 28.670	 2.045   8\\
% }\enesSimulUmono
% % \pgfplotstableread[row sep=\\]{
% % ASR-BLEU	StartOffset	Frames\\
% % 26.165	0.195    2\\
% % 26.387  0.455    4\\
% % 26.472	0.936    6\\
% % 26.585	1.177    8\\
% % }\ennlSimulU-mono
% \pgfplotstableread[row sep=\\]{
% ASR-BLEU	StartOffset	Frames\\
% 26.507	0.103        2\\
% 26.343  0.529        4\\
% 26.872	1.121        6\\
% 26.430	1.492        8\\
% }\enptSimulUmono
% \pgfplotstableread[row sep=\\]{
% ASR-BLEU	StartOffset	Frames\\
% 13.068	1.490    2\\
% 12.91   1.860    4\\
% 12.356	2.101    6\\
% 12.492	2.582    8\\
% }\enruSimulUmono
% % \pgfplotstableread[row sep=\\]{
% % ASR-BLEU	StartOffset	Frames\\
% % 21.305	1.120        2\\
% % 21.909  1.527        4\\
% % 21.628	1.971        6\\
% % 21.720	2.342        8\\
% % }\enroSimulUmono

\begin{figure*}[!t]
\begin{minipage}{0.32\textwidth}
\centering
\begin{tikzpicture}
    \begin{axis}[
            ymajorgrids=true,
            ymin=0,
            ymax=25,
            ytick={5, 10, 15, 20, 25, 30},
            xmin=0,
            xmax=3.5,
            ylabel=ASR-BLEU,
            xlabel=Start Offset (s),
            width=5cm,
            height=4cm,
            title={(a) en-de},
            xlabel style={font=\small},
            ylabel style={font=\small},
            tick label style={font=\footnotesize},
            title style={font=\small, yshift=-1ex},
        ]
        \addplot[color=colStreamAttXTTS,     mark=*] table[x=StartOffset, y=ASR-BLEU]{\endeStreamAttXTTS};
        \addplot[color=colStreamAttSeamless, mark=*] table[x=StartOffset, y=ASR-BLEU]{\endeStreamAttSeamless};
        \addplot[color=colLASeamless, mark=*] table[x=StartOffset, y=ASR-BLEU]{\endeLASeamless};
        \addplot[color=colLAXTTS,     mark=*] table[x=StartOffset, y=ASR-BLEU]{\endeLAXTTS};
        \addplot[color=colSimulU,      mark=*] table[x=StartOffset, y=ASR-BLEU]{\endeSimulU};
    \end{axis}
\end{tikzpicture}
\end{minipage}
\begin{minipage}{0.32\textwidth}
\centering
\begin{tikzpicture}
    \begin{axis}[
            ymajorgrids=true,
            ymin=5,
            ymax=40,
            ytick={5, 10, 15, 20, 25, 30, 35, 40},
            xmin=0,
            xmax=3.5,
            ylabel=ASR-BLEU,
            xlabel=Start Offset (s),
            width=5cm,
            height=4cm,
            title={(b) en-fr},
            xlabel style={font=\small},
            ylabel style={font=\small},
            tick label style={font=\footnotesize},
            title style={font=\small, yshift=-1ex},
        ]
        \addplot[color=colSimulU,      mark=*] table[x=StartOffset, y=ASR-BLEU]{\enfrSimulU};
        \addplot[color=colStreamAttXTTS,     mark=*] table[x=StartOffset, y=ASR-BLEU]{\enfrStreamAttXTTS};
        \addplot[color=colStreamAttSeamless, mark=*] table[x=StartOffset, y=ASR-BLEU]{\enfrStreamAttSeamless};
        \addplot[color=colLASeamless, mark=*] table[x=StartOffset, y=ASR-BLEU]{\enfrLASeamless};
        \addplot[color=colLAXTTS,     mark=*] table[x=StartOffset, y=ASR-BLEU]{\enfrLAXTTS};
    \end{axis}
\end{tikzpicture}
\end{minipage}
\begin{minipage}{0.32\textwidth}
\centering
\begin{tikzpicture}
    \begin{axis}[
            ymajorgrids=true,
            ymin=0,
            ymax=30,
            ytick={5, 10, 15, 20, 25, 30},
            xmin=0,
            xmax=4,
            ylabel=ASR-BLEU,
            xlabel=Start Offset (s),
            width=5cm,
            height=4cm,
            title={(c) en-it},
            xlabel style={font=\small},
            ylabel style={font=\small},
            tick label style={font=\footnotesize},
            title style={font=\small, yshift=-1ex},
        ]
        \addplot[color=colSimulU,      mark=*] table[x=StartOffset, y=ASR-BLEU]{\enitSimulU};
        % \addplot[color=colSimulUmono,  mark=o] table[x=StartOffset, y=ASR-BLEU]{\enitSimulUmono};
        \addplot[color=colStreamAttXTTS,     mark=*] table[x=StartOffset, y=ASR-BLEU]{\enitStreamAttXTTS};
        \addplot[color=colStreamAttSeamless, mark=*] table[x=StartOffset, y=ASR-BLEU]{\enitStreamAttSeamless};
        \addplot[color=colLASeamless, mark=*] table[x=StartOffset, y=ASR-BLEU]{\enitLASeamless};
        \addplot[color=colLAXTTS,     mark=*] table[x=StartOffset, y=ASR-BLEU]{\enitLAXTTS};
    \end{axis}
\end{tikzpicture}
\end{minipage}

% \vspace{-0.5cm}
% Row 2
\begin{minipage}{0.32\textwidth}
\centering
\begin{tikzpicture}
    \begin{axis}[
            ymajorgrids=true,
            ymin=0,
            ymax=35,
            ytick={5, 10, 15, 20, 25, 30, 35},
            xmin=0,
            xmax=3.5,
            ylabel=ASR-BLEU,
            xlabel=Start Offset (s),
            width=5cm,
            height=4cm,
            title={(d) en-es},
            xlabel style={font=\small},
            ylabel style={font=\small},
            tick label style={font=\footnotesize},
            title style={font=\small, yshift=-1ex},
        ]
        \addplot[color=colSimulU,      mark=*] table[x=StartOffset, y=ASR-BLEU]{\enesSimulU};
        \addplot[color=colStreamAttXTTS,     mark=*] table[x=StartOffset, y=ASR-BLEU]{\enesStreamAttXTTS};
        \addplot[color=colStreamAttSeamless, mark=*] table[x=StartOffset, y=ASR-BLEU]{\enesStreamAttSeamless};
        \addplot[color=colLASeamless, mark=*] table[x=StartOffset, y=ASR-BLEU]{\enesLASeamless};
        \addplot[color=colLAXTTS,     mark=*] table[x=StartOffset, y=ASR-BLEU]{\enesLAXTTS};
    \end{axis}
\end{tikzpicture}
\end{minipage}
\begin{minipage}{0.32\textwidth}
\centering
\begin{tikzpicture}
    \begin{axis}[
            ymajorgrids=true,
            ymin=0,
            ymax=30,
            ytick={5, 10, 15, 20, 25, 30,35},
            xmin=-0.1,
            xmax=3,
            ylabel=ASR-BLEU,
            xlabel=Start Offset (s),
            width=5cm,
            height=4cm,
            title={(e) en-pt},
            xlabel style={font=\small},
            ylabel style={font=\small},
            tick label style={font=\footnotesize},
            title style={font=\small, yshift=-1ex},
        ]
        \addplot[color=colSimulU,      mark=*] table[x=StartOffset, y=ASR-BLEU]{\enptSimulU};
        \addplot[color=colStreamAttXTTS,     mark=*] table[x=StartOffset, y=ASR-BLEU]{\enptStreamAttXTTS};
        \addplot[color=colStreamAttSeamless, mark=*] table[x=StartOffset, y=ASR-BLEU]{\enptStreamAttSeamless};
        \addplot[color=colLASeamless, mark=*] table[x=StartOffset, y=ASR-BLEU]{\enptLASeamless};
        \addplot[color=colLAXTTS,     mark=*] table[x=StartOffset, y=ASR-BLEU]{\enptLAXTTS};
    \end{axis}
\end{tikzpicture}
\end{minipage}
\begin{minipage}{0.32\textwidth}
\centering
\begin{tikzpicture}
    \begin{axis}[
            ymajorgrids=true,
            ymin=0,
            ymax=20,
            ytick={5, 10, 15, 20, 25, 30},
            xmin=0,
            xmax=3.5,
            ylabel=ASR-BLEU,
            xlabel=Start Offset (s),
            width=5cm,
            height=4cm,
            title={(f) en-ru},
            xlabel style={font=\small},
            ylabel style={font=\small},
            tick label style={font=\footnotesize},
            title style={font=\small, yshift=-1ex},
        ]
        \addplot[color=colSimulU,      mark=*] table[x=StartOffset, y=ASR-BLEU]{\enruSimulU};
        \addplot[color=colStreamAttXTTS,     mark=*] table[x=StartOffset, y=ASR-BLEU]{\enruStreamAttXTTS};
        \addplot[color=colStreamAttSeamless, mark=*] table[x=StartOffset, y=ASR-BLEU]{\enruStreamAttSeamless};
        \addplot[color=colLASeamless, mark=*] table[x=StartOffset, y=ASR-BLEU]{\enruLASeamless};
        \addplot[color=colLAXTTS,     mark=*] table[x=StartOffset, y=ASR-BLEU]{\enruLAXTTS};
    \end{axis}
\end{tikzpicture}
\end{minipage}

\vspace{-0.5cm}
% Row 3 - two plots centered
\hfill
\begin{center}
\begin{minipage}{0.32\textwidth}
\centering
\begin{tikzpicture}
    \begin{axis}[
            ymajorgrids=true,
            ymin=0,
            ymax=25,
            ytick={5, 10, 15, 20, 25, 30},
            xmin=0,
            xmax=3,
            ylabel=ASR-BLEU,
            xlabel=Start Offset (s),
            width=5cm,
            height=4cm,
            title={(g) en-ro},
            xlabel style={font=\small},
            ylabel style={font=\small},
            tick label style={font=\footnotesize},
            title style={font=\small, yshift=-1ex},
        ]
        \addplot[color=colSimulU,      mark=*] table[x=StartOffset, y=ASR-BLEU]{\enroSimulU};
        % \addplot[color=colStreamAttXTTS,     mark=*] table[x=StartOffset, y=ASR-BLEU]{\enroStreamAttXTTS};
        \addplot[color=colStreamAttSeamless, mark=*] table[x=StartOffset, y=ASR-BLEU]{\enroStreamAttSeamless};
        \addplot[color=colLASeamless, mark=*] table[x=StartOffset, y=ASR-BLEU]{\enroLASeamless};
        % \addplot[color=colLAXTTS,     mark=*] table[x=StartOffset, y=ASR-BLEU]{\enroLAXTTS};
    \end{axis}
\end{tikzpicture}
\end{minipage}
\begin{minipage}{0.32\textwidth}
\centering
\begin{tikzpicture}
    \begin{axis}[
            ymajorgrids=true,
            ymin=0,
            ymax=35,
            ytick={5, 10, 15, 20, 25, 30, 35},
            xmin=-0.1,
            xmax=2.5,
            ylabel=ASR-BLEU,
            xlabel=Start Offset (s),
            width=5cm,
            height=4cm,
            title={(h) en-nl},
            xlabel style={font=\small},
            ylabel style={font=\small},
            tick label style={font=\footnotesize},
            title style={font=\small, yshift=-1ex},
        ]
        \addplot[color=colSimulU,      mark=*] table[x=StartOffset, y=ASR-BLEU]{\ennlSimulU};
        \addplot[color=colStreamAttXTTS,     mark=*] table[x=StartOffset, y=ASR-BLEU]{\ennlStreamAttXTTS};
        \addplot[color=colStreamAttSeamless, mark=*] table[x=StartOffset, y=ASR-BLEU]{\ennlStreamAttSeamless};
        \addplot[color=colLASeamless, mark=*] table[x=StartOffset, y=ASR-BLEU]{\ennlLASeamless};
        \addplot[color=colLAXTTS,     mark=*] table[x=StartOffset, y=ASR-BLEU]{\ennlLAXTTS};
    \end{axis}
\end{tikzpicture}
\end{minipage}
\end{center}

\begin{center}
\begin{tikzpicture}
\begin{axis}[
        hide axis,
        xmin=0, xmax=1, ymin=0, ymax=1,
        width=4cm, height=2cm,
        legend style={
            draw=black,
            at={(0.5,0.5)},
            anchor=center,
            legend columns=-1,
            font=\small,
            column sep=0.5ex,
            /tikz/every even column/.append style={column sep=1.5ex},
        },
        legend cell align={left},
    ]
    \addplot[color=colSimulU,      mark=*] coordinates {(0,0)};
    \addplot[color=colStreamAttXTTS,     mark=*] coordinates {(0,0)};
    \addplot[color=colStreamAttSeamless, mark=*] coordinates {(0,0)};
    \addplot[color=colLASeamless, mark=*] coordinates {(0,0)};
    \addplot[color=colLAXTTS,     mark=*] coordinates {(0,0)};
    \legend{SimulU, StreamAtt+XTTS-v2, StreamAtt+Seam.TTS, LA+Seam.TTS, LA+XTTS-v2}
\end{axis}
\end{tikzpicture}
\end{center}
\vspace{-0.3cm}
\caption{ASR-BLEU and Start Offset scores across different systems for each language pair of MuST-C v1 tst-COMMON.}
\label{fig:system_comparison}
\end{figure*}
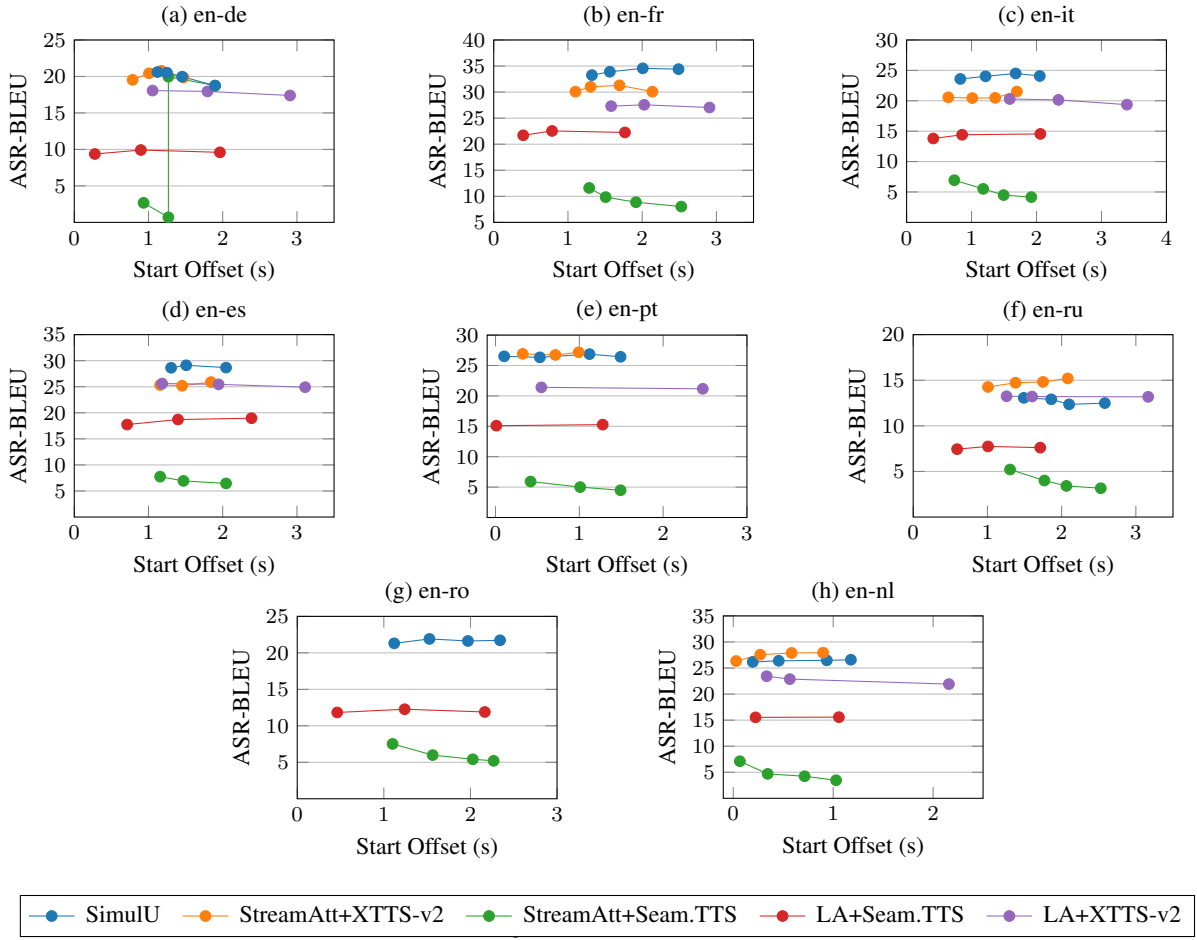

\begin{table}[t]
    \centering
    \addtolength{\tabcolsep}{-4pt}
    \resizebox{\linewidth}{!}{
    \begin{tabular}{llcccccccc}
    \toprule
    \multirow{1}{*}{\textbf{TTS sys.}} & \multirow{1}{*}{\textbf{Metric}} & \texttt{de} & \texttt{fr} & \texttt{it} & \texttt{nl} & \texttt{pt} & \texttt{es} & \texttt{ru} & \texttt{ro} \\
    %&      & \texttt{2640} & \texttt{2631} & \texttt{2509} & \texttt{2594} & \texttt{2444} & \texttt{2500} & \texttt{2467} & \texttt{2490} \\
    \midrule
    \multirow{2}{*}{\textit{Seam.TTS}}
        & WER (\%)  & 41.77 & 35.28 & 39.03 & 40.81 & 43.29 & 34.59 & 46.89 & 40.15 \\
       & CER (\%)  & 31.76 & 28.75 & 31.29 & 31.33 & 34.41 & 29.14 & 35.56 & 26.29 \\
        & UTMOS     & 2.82  & 2.60  & 3.63 & 3.50 & 3.87 & 3.60  & 3.32  & 3.83  \\
        & ($\pm$ std) & (0.32) & (0.35) & (0.34) & (0.27) & (0.33) & (0.33) & (0.41) & (0.29) \\
        
    \midrule
    \multirow{2}{*}{\textit{XTTS-v2}}
        & WER (\%)  & 9.789 & 14.02 & 10.25 & 10.19 & 6.24 & 3.93 & 9.78 & ---\\
        & CER (\%)  & 7.70 & 10.60 & 7.85  & 7.00  & 3.17 & 2.38 & 6.49 & ---\\
        & UTMOS     & 2.77 & 2.53 & 2.71 & 2.94 & 2.83 & 2.66 & 2.87 & ---\\
        & ($\pm$ std) & (0.40) & (0.40) & (0.38) & (0.34) & (0.37) & (0.36) & (0.33) & ---\\
    \bottomrule \\
    \end{tabular}}
    \caption{TTS results for the Seam.TTS and XTTS-v2 systems.}
    \label{tab:wer_cer_tts_systems}
    \vspace{-3em}
\end{table}

Figure~\ref{fig:system_comparison} reports test-set performance of the proposed method, SimulU, alongside the previously defined strong cascade systems (Section \ref{subsec:models}). 
Detailed results for each language pair suggest that SimulU consistently achieves the highest ASR-BLEU scores across six language directions (de, fr, it, es, pt, and ro), while maintaining competitive performance in the other ones (ru and nl), with always a reasonable latency (between 1 and 2 seconds in most cases) as measured by the start offset.\footnote{Limits of acceptability have been set at $\sim$2 seconds for the \textit{ear-voice span} depending on different conditions and language pairs \cite{CHMIEL_2017}.}
As shown, both SimulU and StreamAtt+XTTS-v2 outperform both LA-based systems by a large margin (at least, 4-5 ASR-BLEU points at the same latency), particularly in fr, pt, ro, and nl. The LA-based cascades span a wide range of start offset (often between 1 and 3 seconds), and increasing the speech segment size--hence, the available context--always yields similar performance (e.g., in fr, it, pt, ro, nl).
Replacing the strong XTTS-v2 model (as attested by results in Table~\ref{tab:wer_cer_tts_systems}) with Seam.TTS, as the TTS component in the cascades, leads to substantial quality degradation, with ASR-BLEU scores oscillating between 5 and 10 with the StreamAtt policy, indicating near-complete failure, and between 10 and 15 with the LA policy. 
A manual inspection suggests that this degradation largely stems from Seam.TTS's sensitivity to limited context: when conditioned on partial sentences rather than complete ones (as is typical in SimulS2ST settings), its synthesis quality deteriorates markedly, while it is not the case for XTTS-v2.

We further examine the end-offset latency, defined as the time delay between the end of the input speech stream and the generation of the final speech output by the system. 
Table \ref{tab:endoffset_systems} reports the results for the two best-performing systems, SimulU and StreamAtt+XTTS-v2 (SimulU and StreamAtt+Seam.TTS for Romanian). 
Considering the average performance across four cut-off frame settings for SimulU and across segment step configurations for StreamAtt+XTTS-v2, we observe that systems with comparable overall performance exhibit similar end-offset latency in de and fr. 
However, for the other languages, SimulU achieves lower end-offset latency values. 
Furthermore, the standard deviation under the SimulU policy is smaller, indicating more stable latency behavior.

All in all, we can conclude that SimulU achieves the best quality overall across the analyzed languages while maintaining the latency between 1 and 2 seconds.

\begin{table}[t]
    \centering
    \addtolength{\tabcolsep}{-1pt}
    \resizebox{\linewidth}{!}{
    \begin{tabular}{ccccccccc}
    \toprule
    \multirow{1}{*}{\textbf{System}} & \texttt{de} & \texttt{fr} & \texttt{it} & \texttt{nl} & \texttt{pt} & \texttt{es} & \texttt{ru} & \texttt{ro} \\
    \midrule
    \multirow{2}{*}{\textit{SimulU}}
        & 247   & 224   & 100 & 82  & 146  & 106 & 106  & 34  \\
        & (137) & (132) & (6) & (9) & (6) & (5) & (13) & (2) \\
        
    \midrule
    \multirow{1}{*}{\textit{Top}}
         &  246  &  224  & 262  & 40  & 251  & 217  &  68 & 58  \\
     \multirow{1}{*}{\textit{Cascade}}   & (137) & (132) & (122) & (44) & (123) & (124) & (62) & (52) \\
    \bottomrule \\
    \end{tabular}}
    \caption{End Offset in milliseconds (mean and std) for SimulU and the top cascade for each language averaged across cutoff frame $f$ and speech segment size, respectively.}
    \label{tab:endoffset_systems}
\end{table}

\section{Conclusions}
In this work, we propose SimulU, the first training-free long-form simultaneous speech-to-speech translation policy that exploits the internal cross-attention of pre-trained end-to-end models to regulate both input history and output generation dynamically.
We evaluated its performance across eight language pairs against strong cascade systems combining top-performing existing solutions for streaming translation and TTS systems. 
In these settings, SimulU yields strong results even when compared with state-of-the-art cascades, demonstrating the effectiveness of the proposed approach without requiring any additional training, therefore offering a practical and competitive solution for real-world simultaneous speech translation.

\clearpage

% \section{Generative AI Disclosure}
% Portions of this manuscript were reviewed and refined using generative AI tools. Specifically, AI assistance was used to improve grammar, enhance academic writing style, and refine table formatting. All technical content, experimental design, analysis, and conclusions were developed and verified by the authors.

\section{Acknowledgmements}
This work has received funding from the European Union’s Horizon Europe programme under grant agreement No. 101213369 (DVPS).
The research was conducted during an internship at FBK, which was facilitated by MBZUAI Career Services, whose support we gratefully acknowledge.
We also thank Dr. Hanan Aldarmaki for supporting the main author through valuable feedback and guidance.

\bibliographystyle{IEEEtran}
\bibliography{mybib}

\end{document}